\pdfoutput=1
\documentclass[twocolumn,showpacs,preprintnumbers,amsmath,amssymb,prb]{revtex4}

\usepackage{graphicx}
\usepackage{dcolumn}
\usepackage{bm}
\usepackage{graphicx}
\usepackage[3D]{movie15}
\usepackage[colorlinks=false]{hyperref}

\newcommand{\suml}{\sum\limits}

\newcommand{\R}{\mathbb{R}}
\newcommand{\cH}{\mathcal{H}}

\newcommand{\conj}[1]{{#1}^{\star}}
\newcommand{\deba}{\partial}

\begin{document}

\title{Multiband Hamiltonians of the Luttinger-Kohn Theory and Ellipticity Requirements}

\author{Dmytro Sytnyk}
 \email{sytnikd@gmail.com}
 \affiliation{%
$M^{2}NeT$ Laboratory,
Wilfrid Laurier University,
75 University Avenue West,
Waterloo, ON,
Canada, N2L 3C5.
}%

\author{Sunil Patil}%
 \email{spatil@wlu.ca}
 \affiliation{%
$M^{2}NeT$ Laboratory,
Wilfrid Laurier University,
75 University Avenue West,
Waterloo, ON,
Canada, N2L 3C5.
}%
\author{Roderick Melnik}
 \email{rmelnik@wlu.ca}
 \affiliation{%
$M^{2}NeT$ Laboratory,
Wilfrid Laurier University,
75 University Avenue West,
Waterloo, ON,
Canada, N2L 3C5.
}%
\begin{abstract}
Modern applications require a robust and theoretically strong tool for the realistic modeling of electronic states in low dimensional nanostructures.  The $k \cdot p$ theory has fruitfully served this role for the long time since its creation. During last two decades several problems have been detected in connection with the application of the $k \cdot p$ approach to such nanostructures.
These problems are closely related to the violation of the ellipticity conditions  for the underlying system, the fact that until recently has been largely overlooked.
We demonstrate that in many cases the models derived by a formal application of the Luttinger-Kohn theory fail to satisfy the ellipticity requirements.
The detailed analysis, presented here on an example of the $6 \times 6$ Hamiltonians, shows that this failure has a strong impact on the physically important properties conventionally studied with these models.
\end{abstract}


\maketitle

The effective mass theory is one of the fundamental parts in the physics of nanostructures.
This theory allows us to get theoretical insight into the electronic properties of dominating bands near the
extremum points \cite{BP1, ChuangB_95}.  Futhermore, the theory establishes a
robust computational framework for simulating observable quantum-mechanical states and corresponding
energies in the low-dimensional systems including quantum wells, wires, nanodots.
In the original Luttinger--Kohn work \cite{Lutt_55}
authors applied the theory to the Schr\"odinger equation perturbed by a smooth potential and constructed a
representation for valence bands Hamiltonian near high symmetry point $\Gamma$ of the first
Brillouin zone in bulk Zinc-Blende (ZB) crystals with large fundamental band gap.
Soon after that, Kane showed how to expand the model to the narrow gap materials  such as InSb and Ge
for instance, where one can also account for the influence of the conduction bands \cite{Kane1957}.
One of the advantages of the $k\cdot p $ theory is in its universality. Indeed, the theory had also been extended to cover Wurtzite (WZ) type of crystals,
materials with inclusions, heterostructure materials and superlatices \cite{Xia91}.
Another advantage of the effective mass theory that has recently been explored in some details in \onlinecite{Patil2009}
is its flexibility,
as one can easily adjust the
models to include additional effects like strain, piezoelectricity, magnetic field,
and respective nonlinear effects.
These inbuilt multi-scale effects are  crucial for such applications as light-emission diodes,
lasers, high precision sensors, photo-galvanic elements, hybrid bio-nanodevices, and many others \cite{Abajo2007}.

For a wide range of applications these models have provided good, computationally feasible and
efficient approximations that agree well with experimental results \cite{ChuangB_95, ChuangB_09}.
However, for some types of crystal materials band structure calculations based on such multiband models lead
to the solutions with unphysical properties \cite{Smith_sp_86, Szmulowicz_96} or so called spurious solutions \cite{foreman_sp_07, Yang_sp_05, Melnik2009}.

As a result, there have been various attempts to explain the origin of the spurious solutions and
develop some reliable procedures on how to avoid them \cite{Veprek07}.
 These approaches rely on three main ideas: (a) to modify the original Hamiltonian and remove the terms
 responsible for the spurious solutions \cite{Kolokolov_sp_03, Yang_sp_05}, (b) to
 change band-structure parameters \cite{Eppenga_sp_87, foreman_sp_07}, and (c) to identify and exclude physically
 inadequate observable states \cite{Kisin_sp98}.
 All mentioned approaches suffer from the common weakness -- the lack
 of clear justification of the underlying theoretical procedure and thus from limitations in their applicability
 \cite{Veprek07, Veprek08}.

In this work we show that spurious solutions are not just a reason but rather an imminent consequence of
another fundamental problem in applications of the classical theory -- the non-ellipticity of
the multiband Hamiltonian. The widely adopted effective mass approximations of the
original Schr\"odinger elliptic Hamiltonian turns out to be non-elliptic for a broad class of known material
parameters (cf. Table \ref{tabZBDRpar}). This fact leads to the following consequence: since any qualitative
approximation methods must preserve the
topological structure of the spectrum of a general linear operator, and as an implication symmetric properties
in case of Schr\"odinger Hamiltonian, this evident discrepancy signifies
the mathematical invalidity of performed approximation procedures for those materials.
Among these mutiband Hamiltonians there are a number of widely used $6 \times 6$ and $8\times 8$ Hamiltonians.

The paper is organized as follows.
First, we revise basic properties of the original Schr\"odinger equation,
and outline the mathematical model of the electronic band-structure problem.

%
 Next, we derive the exact constraints on the material parameters for typical $6\times 6$
 Hamiltonians in ZB \cite{Lutt_55, Kane1957} and WZ \cite{BP1} materials.
 Direct calculations using conventional Luttinger parameters (e.g. \onlinecite{LB1, Madelung2004}) show that
 such parameters (e.g. Table \ref{tabZBDRpar}) for many important semiconductor materials entail the violation of
 ellipticity requirements. As a result, the corresponding bandstructure model for them potentially susceptible
 to unphysical solutions,  even in the bulk case, and therefore ought to be modified.
 Moreover, the corresponding time dependent Schr\"odinger equation loses the fundamental property
 of state conservation \cite{Landau_qm1982}.


The material properties (such as fundamental band-gaps and spin-orbit splitting energies)
obtained experimentally, represent real phenomena, whereas models based on finite bands Hamiltonians
are meant to approximate them.
The very last step in such approximation schemes \cite{Eppenga_sp_87} enables us to calculate interband corrections
to the main part of the Hamiltonians with help of perturbation theory \cite{Lutt_55, Lutt_56, BP1}.
However, this last step lacks a rigorous theoretical foundation as it does not guaranty the convergence of the perturbation expansion.
The result is that the derived Hamiltonian, although directly based on experimental parameters
(column 4, \ref{tabZBDRpar}), represents a totally different mathematical object compared
to its origin.
The physical evidences, to support this claim have been already known for
GaAs \cite{Pfeffer1990} and recently been reported for Si \cite{Dorozhkin_08}.


We start with the Schr\"odinger equation with the potential $V(x)$
\begin{equation}\label{SchrodEqs}
H_0\psi(x) \equiv \frac{\mathbf{p}^2}{2m^\star}\psi(x)+(V(x)+E^\star_0)\psi(x)=E\psi(x),
\end{equation}
where $E^\star_0$ is the band edge energy of the system, $\mathbf{p}$ is a momentum operator,
$x \in \Omega \subset {\mathbb{C}}^n, n\leq3$, $m^\star$
is a piecewise-constant effective mass parameter.
In $\Omega$ we supplement (\ref{SchrodEqs}) by usual Dirichlet conditions on the boundary
$S \equiv \partial \Omega$
\begin{equation}
\label{SEBC}
\psi(x) = 0, \quad x \in S.
\end{equation}
Our major focus is on the heterostructure case, where the difference in the effective mass and probability
current conservation imposes the discontinuity of $\nabla{\psi(x)}$ on the interface between materials
(assuming that the basis Bloch functions are the same in all constituents \cite{Burt_99}).
In this case an appropriate choice of the functional space for coefficients from (\ref{SchrodEqs})
is the Hilbert space of distributions  with the compact support $\cH^1(\Omega)$ \cite{Hormander1, Hormander2}.

Then, the operator $H_0$ with the domain of definition $D(H_0) \subset \cH^3(\Omega)$ is a symmetric operator with an existing self-adjoint extension, hence it conserves the probability current \cite{Dirac81}.
General theory of elliptic partial differential equations characterizes the problem (\ref{SchrodEqs})
in terms of the corresponding Fredholm type theorems
(e.g. \onlinecite{Egorov11998} 
),
so that the problem (\ref{SchrodEqs}), (\ref{SEBC}) has a countable set of eigenvalues $E_i$,
smallest eigenvalue $E_0$ is simple and corresponding eigenstate $\psi_0$ is of constant
sign in $\Omega$, furthermore
$$
E_i \in [\rho_0, \infty),\  \rho_0 = E^\star_0 + \min\{0,\inf_{\Omega}{V(x)} \},\quad i=0,\ldots n,
$$
where V(x) is real. 

If $V(x)$ is a gently varying function over the unit cell in the sense of \onlinecite{Lutt_55} the
original operator
$H_0$ can be approximated by another operator $H$ (using Bloch theorem), determined by the projection
$P$ of $H_0$ on the considered eigenspace and
Lowding perturbation theory \cite{Lutt_55, BP1}.
The last step in this approximation procedure accounts for the influence of the elements from the
space complement to the egeinspace by the formula
\begin{equation}\label{eq_PO}
  H = P H_0 + \suml_{i=1}^r \lambda^r H^{(r)}
\end{equation}
up to the order $r$. Setting $\lambda=1$ leads one to the final approximation, under the assumption that the series (\ref{eq_PO})
is convergent for such $\lambda$.
Despite a wide applicability of such approximations,
the intrinsic ellipticity requirements for the realizations of $H$ have not been explicitly verified
in a systematic manner (see \onlinecite{Lutt_55, Lutt_56, Kane1957, BP1, ChuangB_95}, as well as more recent works).

As an example, let us consider two classical Hamiltonians for ZB \cite{Lutt_55} and WZ \cite{BP1}
type of materials with more scrutiny.
In what follows we use the parameter notation identical to the works where corresponding Hamiltonians were
obtained. When necessary, the parameters will be converted from one notation to another by using the
formulas  from [{p. 82}, \onlinecite{Cardona_05}].

First consider the Luttinger-Kohn (LK) Hamiltonian from \onlinecite{Lutt_55}
\begin{widetext}
\begin{eqnarray*}\label{H_LK}
H^{LK} \equiv \left(
\begin{array}{ccccccc}
\frac{1}{2} P & L & M & 0 & i \frac{1}{\sqrt{2}} L & -i \sqrt{2} M\\
\conj{L} & \frac{1}{6} P+\frac{2}{3} Q & 0 & M & -\frac{i}{3\sqrt{2}} (P-2 Q) & i \sqrt{\frac{3}{2}} L\\
\conj{M} & 0 & \frac{1}{6} P+\frac{2}{3} Q & -L & -i \sqrt{\frac{3}{2}} \conj{L} & -\frac{i}{3\sqrt{2}}
(P-2 Q)\\
0 & \conj{M} & -\conj{L} & \frac{1}{2} P & -i \sqrt{2} \conj{M} & -\frac{i}{\sqrt{2}} \conj{L}\\
-\frac{i}{\sqrt{2}} \conj{L} & \frac{i}{3\sqrt{2}} (P-2 Q) & i \sqrt{\frac{3}{2}} L & i \sqrt{2}
M & \frac{1}{3} (P+Q) & 0\\
i \sqrt{2} \conj{M} & -i \sqrt{\frac{3}{2}} \conj{L} & \frac{i}{3\sqrt{2}} (P-2 Q)
& \frac{i}{\sqrt{2}} L & 0 & \frac{1}{3} (P+Q)\\
\end{array}
\right),\\
\end{eqnarray*}
\end{widetext}
where each of the $P, Q, L, M$ is a second order position dependent differential operator or equivalently second order polynomial in the momentum representation \cite{Lutt_55}:
$$
  \begin{array}{c}
    P=(A+B) k_{+}k_{-}+2 B k_z^2,\quad
    Q=B k_{+}k_{-}+A k_z^2,\\[11pt]
    M=\frac{1}{2\sqrt{3}} ((A-B) (k_x^2-k_y^2)-2 i C k_x k_y),\\[11pt]
    L=-i \frac{C}{2\sqrt{3}} k_{-} k_z,\quad
    k_{\pm}=k_x\pm i k_y.
  \end{array}
$$
Our aim is to check the type (elliptic, hyperbolic or essentially hyperbolic) of the $H^{LK}$ as a
partial-differential operator (PDO) on $\cH^1(\Omega)$ as we know that the given Schr\"odinger operator
from (\ref{SchrodEqs}) is elliptic.
Only the second order derivative terms are playing the dominant role in the following analysis because
contributions from the terms linear in $k$ as well as from the potential, are bounded in the domain $D(H^{LK})$ \cite{Hormander2}.
It means that the results for more complicated physical models with potential contributions from
additional fields (e.g. strain, magnetic field, etc.) will stay the same as for the original $H^{LK}$, analyzed here.
The fact that the Hamiltonian is a linear operator guarantees that it is also true for any
other representation of $H^{LK}$ obtained by linear (basis) transformations.

In a more general sense, for any $m$--dimensional matrix PDO $H=\{h_{ij}\}_{i,j =1}^{m}$,
where each element $h_{ij}$ is a second order one dimensional PDO \cite{Egorov11998, Hormander1}
\begin{equation}\label{pdo}
h_{ij} = \suml_{k,l =0}^n h_{ij}^{kl} \frac{\deba^2}{\deba x_k \deba x_l},
\end{equation}
the associated quadratic form is defined by
\begin{equation}\label{QF}
G(\xi_1, ..., \xi_{nm}) = v M v^T, \quad v = \left(\xi_1, \ldots , \xi_{nm}\right),
\end{equation}
where $M$ is an $mn \times mn$ matrix composed from the elements $h_{ij}^{kl}$.
The $\mathrm{k\cdot p}$ Hamiltonians in $\R^3$ are a special case of (\ref{pdo})-(\ref{QF}) with the real eigenvalues $\alpha_i$ and $n=3$
(e. g. \onlinecite{Veprek07}).

Using these notations, the procedure of obtaining the ellipticity condition for $H$ reduces to the question about
the sign of $\lambda_i$ for the associated $M$.
More precisely, the matrix differential operator $H$ will be elliptic if and only if all eigenvalues
of the corresponding Hermitian $M$ will have the same sign \cite{Hormander2, Egorov11998}.

In general, it is a challenging task to calculate the eigenvalues of $M$ explicitly, even for such small as
$3\times 3$ Hamiltonians, but we recall that we deal here with usually sparse, band structure operators.

Taking into account the fact that the sequence of eigenenergies of $H_0$ is semi-bounded, for an approximation $H^{LK}$ (in the momentum representation), we obtain
\begin{equation}\label{eqineig}
\lambda_i<0, \quad \forall \ i=0,1,\ldots, nm.
\end{equation}
The last constrains guarantee the ellipticity (in strong sense  \cite{Hormander1}) of Hamiltonian $H$.
The operator $H$ possess a self-adjoint extension in $D(H)\subset \cH^{k+2}(\Omega)$, $k>0$,
%
provided that the domain $\Omega$ is sufficiently smooth (piecewise Lipschitz). 
Then it can be extended to a Hermitian operator by closure in the norm [p. 113, \onlinecite{Egorov11998}] or via Lax-Miligram procedure \cite{Hormander2}.
From the physical point of view the smoothness characteristics of $D(H)$ fulfill the natural assumption of quantum theory that the state of the system must be
a continuous function of spatial variables even when some coefficients of $H$ have finite jumps \footnote{In this case the smoothness coefficient $k$ in the definition of domain and image of $H$ is $k=1$.}
like in the heterostructures consisting of different materials \cite{ChuangB_95, Burt_99}.

%
\begin{table}
  \centering
  \caption{The material parameters for ZB type of materials,
  $d$ --  distance from the point $(A,B,C)$ to the ellipticity region $\Lambda_{-}(H^{LK})$} \label{tabZBDRpar}
\begin{ruledtabular}
  \begin{tabular}{r|ccccr|cccc}
 El. & $A$& $B$& $C$& $d$& El. & $A$& $B$& $C$& $d$\\
 \hline
AlAs\footnotemark[2]&
-7.5& -2& 8.4& 1.97&
Ge\footnotemark[5]\footnotemark[2]&
-30& -4.6& 33& 10.64\\
AlP\footnotemark[1]&
-3.8& -3.4& 7.2& 0.92&
Ge\footnotemark[4]\footnotemark[5]&
 -30& -4.6& 33& 10.64\\
AlP\footnotemark[5]&
-3.7& -3.4& 7.2& 1.01&
Ge\footnotemark[1]&
-30& -4.4& 36& 11.99\\
GaN\footnotemark[3]&
-7.5& -3.8& 6.1& In&
InP\footnotemark[10]&
-11& -1.8& 10& 2.86\\
C\footnotemark[7]&
-4& -3.4& 6.6& 0.42&
InP\footnotemark[10]\footnotemark[5]&
-15& -2.1& 17& 5.72\\
C\footnotemark[8]&
-2.9& -2.3& 3.9& In&
InP\footnotemark[10]&
-9& -3.3& 9.6& 1.34\\
C\footnotemark[8]&
-6& -3.8& 6& In&
InSb\footnotemark[3]&
-100& -4& 96& 39.35\\
C\footnotemark[8]&
-3.1& -1.7& 0.9& In&
InSb\footnotemark[1]&
-100& -3& 96& 40.25\\
GaAs\footnotemark[9]&
-16& -2& 6& 0.89&
Si\footnotemark[11]&
-6& -3.5& 9.6& 1.18\\
GaAs\footnotemark[9]&
-17& -2.2& 6.6& 0.98&
Si\footnotemark[11]&
-5.5& -3.6& 8.4& 0.54\\
GaAs\footnotemark[9]&
-15& -3.1& 17& 4.83&
Si\footnotemark[11]&
-6& -3.4& 8.4& 0.72\\
GaP\footnotemark[12]&
-6& -3& 7.2& 0.54&
SiC\footnotemark[1]&
-2.6& -1.7& 4.2& 0.38\\
GaP\footnotemark[12]\footnotemark[5]&
-8& -2.2& 10& 2.50&
SiC\footnotemark[3]&
-4.8& -1.8& 5.1& 0.67\\
  \end{tabular}
\end{ruledtabular}
\footnotetext[4]{Measured under $T=4.2K$}
\footnotetext[5]{Obtained by extrapolations from $14\times 14$ $\mathbf{k\cdot p}$ model}
\footnotetext[6]{Measured under $T=300K$}
\footnotetext[1]{Ref. \cite{Madelung2004}}
\footnotetext[2]{Ref. \cite{LB1}}
\footnotetext[3]{Ref. \cite{Cardona_05}}
\footnotetext[7]{Most probable value (set 5 from \cite{Madelung2004})}
\footnotetext[8]{Sets 1, 2, 3 from \cite{Willatzen1994}},
\footnotetext[9]{Sets 7 (T=50K), 8 (T=70K), 2 from \cite{LB1}}
\footnotetext[10]{Sets 7 (T=60..300K), 2, 1  from \cite{LB1}}
\footnotetext[11]{Sets 2 (T=1.26K), 3, 6  from \cite{LB1}}
\footnotetext[12]{Sets 4 (T=1.6K), 3, 6  from \cite{LB1}}
\end{table}

\begin{figure}[ht]
  \begin{center}
\includemovie[
        poster,
	toolbar, 
	label=temp,
    text={\includegraphics[width=\linewidth]{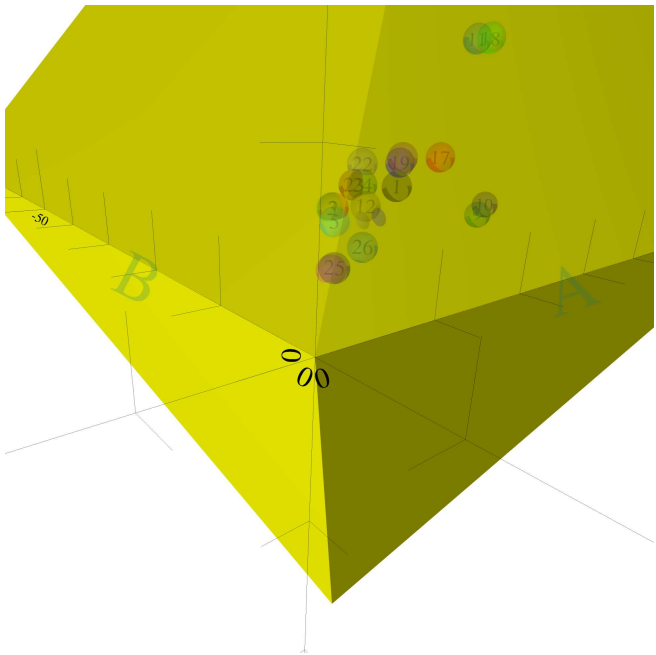}},
        3Droo=0.2233495084617366,
        3Daac=60,
        3Dcoo=0.0000067046325966657605 0.011743400245904922 -0.059044793248176575,
        3Dc2c=0.060472868382930756 -0.054119743406772614 0.9967015981674194,
        3Droll=44.09483043248314,
        3Dlights=Hard,
        3Dviews2=views.3dv,
]{\linewidth}{\linewidth}{ZBLKinABC.u3d}
 \end{center}
  \caption[The region $\Lambda_{-}(H^{LK})$ and the data Table \ref{tabZBDRpar}]{The region $\Lambda_{-}(H^{LK})$ and the material parameters from Table \ref{tabZBDRpar}.(color online)}
  \label{fig:DLK}
\end{figure}

The direct calculation by (\ref{QF}) for $H^{LK}$ ($n=3$, $m=6$) leads us to the $18 \times 18$ matrix $M^{LK}$
with the following distinct eigenvalues:
\begin{equation}\label{eig_LK}
\begin{array}{cc}
\lambda_1 =  - {  \frac {\mathit{C }}{2}}  + \mathit{A },
&    \lambda_{2/3} = ({  \frac {1}{4}}  \pm {  \frac {\sqrt{3}}{4}} )\mathit{C } + \mathit{A },\\[12pt]
\lambda_{4/5} = \pm {  \frac {\mathit{C }}{2}}  + \mathit{B }, &
\lambda_{6/7} =  {\pm  \frac {\mathit{C }}{4}}  + \mathit{B },
\end{array}
\end{equation}
where $\lambda_1$, $\lambda_{2/3}$, $\lambda_{4/5}$ have the multiplicity 2 and $\lambda_{6/7}$  have
the multiplicity 4, respectively,  and $A, B, C$ are usual material parameters \cite{Lutt_55}.
By substituting (\ref{eig_LK}) into (\ref{eqineig}), we receive the system of linear  inequalities with respect to  $A$, $B$ and $C$.
They describe the feasibility region in the $A, B, C$ space, when $H^{LK}$ is an elliptic partial differential operator with the discrete and decreasing sequence of eigenvalues.
One can use similar reasoning to obtain corresponding inequalities for other common representations
of $H^{LK}$ through Luttinger parameters $\gamma_i$ \cite{ChuangB_95, BP1}. Evidently, any solution
of (\ref{eqineig}) for (\ref{eig_LK}) would have a unique corresponding solution in $\gamma_i$ notation.
Note next that the solution domain $\Lambda_{-}(H^{LK})$ of (\ref{eig_LK}), (\ref{eqineig}) is symmetric with
respect to the sign of $C$. It follows from the form of the $H^{LK}$ which is dependent on $|C|$ only.
$\Lambda_{-}(H^{LK})$ comprises an unbounded pyramid in $\R^3$ (cf. Fig. \ref{fig:DLK}) with
the following rays as its edges:
$
\left(-\frac{1}{2}t,-\frac{1}{2}t,-t\right),\  \left(-\left( \frac{1}{4}+\frac{\sqrt{3}}{4}\right)t,
-\frac{1}{2}t, t \right),
$
and $
(-t,0,0),\ \left(0, -\left(\frac{3}{4}+\frac{\sqrt{3}}{4}\right)t, 0 \right),
$
for $t \in [0, \infty)$.
%

Figure \ref{fig:DLK} represents the $H^{LK}$ ellipticity region together with the widely
adopted values of the material parameters for different ZB type materials summarized in Table \ref{tabZBDRpar}.

One can observe that, among all analyzed materials only two
(indicated as "In" in the third column of Table \ref{tabZBDRpar}) have the
admissible sets of parameters.
All other data from Table \ref{tabZBDRpar} yield the condition $\lambda_4>0$. That is why the $H^{LK}$, for the corresponding materials, is not elliptic and may not be even symmetric. Moreover, instead the domain $D(H^{LK})=D(H)\subset \cH^{3}(\Omega)$ we have only
\begin{equation}\label{DLKnonel}
D(H^{LK})=D(H)\subset \cH^{1}(\Omega),
\end{equation}
it means that the solution of (\ref{SchrodEqs}) will have the discontinuities, for the systems with jump discontinues coefficients, which is the case for heterostructure materials, see e.g. \onlinecite{Courant1956}.
The general theory guaranties that, in this case, the interface discontinuities will be observed all through the interior of the active region along the characteristics of $H^{LK}$, which are now shown to exist since the associated $G$ in (\ref{QF}) is of nonconstant sign [p. 153, \onlinecite{Egorov11998}].
Additionally, (\ref{DLKnonel}) and the double degeneracy of $\lambda_4$ (\ref{eig_LK}) would lead to the nonanalytic solution and thus the momentum operator, from (\ref{SchrodEqs}), will be ill-defined (by the embedding theorems, [p. 119, \onlinecite{Egorov11998}]). All the arguments stated before allow us to conclude that the $H^{LK}$ does not provide a sufficiently good approximation, preserving the type of the PDO, for most of the practical data.

  The ellipticity analysis for ZB, can be applied to a WZ $6\times 6$ Hamiltonian \cite{BP1, Suzuki1995} without any changes. Ellipticity conditions that follows from such analysis are, again, linear in parameter variables
\begin{equation}\label{constrWZ}
\begin{array}{cc}
 A_2 + A_4<0,
& A_2\pm (2A_5-2A_4)<0,\\[12pt]
A_2+4 A_5<0, &
A_2+3 A_4-2 A_5<0.
\end{array}
\end{equation}
Namely, $A_i$ are well-known Luttinger-like parameters for WZ \cite{BP1}.
As in the ZB case, each separate inequality has been obtained from (\ref{eqineig}) by substituting every distinct eigenvalue of the matrix $M$ associated with the WZ Hamiltonian.
These conditions are also violated in most of practically important materials, among which, we would like to mention GaN,  AlN and ZnO. Here, the distances to the WZ ellipticity region defined by (\ref{constrWZ}) are approximately equal: 0.804, 0.862, 0.606 for GaN parameter sets \cite{Suzuki1995, Chuang1996, Mireles2000}; 1.132 \cite{Suzuki1995, Chuang1996}, 1.271 \cite{Mireles2000}, 1.01 \cite{JeonWZ96} for AlN; 1.067 \cite{FanZnO2006} for ZnO. The distances have the order of terms in  the unperturbed part of the WZ Hamiltonian (which in the dimensionless Luttinger-like notation equal to 1), and thus are considerably high.

Let us return back to the feasible parameters for two materials C and GaN. For carbon, the parameter values were analyzed in \onlinecite{Reggiani1983}, where authors showed that they don't agree with the Hall effect experimental measurements.
In the same paper the authors suggested another, more consistent (in term of the measurements), set of parameters (row 5, Table \ref{tabZBDRpar}).
Observe, however, that the latter set does not belong to the ellipticity region $\Lambda_{-}(H^{LK})$.
In terms of the distance to $\Lambda_{-}(H^{LK})$, we can also classify other mostly large band gap materials, such as Si, SiC, AlP and GaN, as those belonging to the same group.
For GaN we have the set of $A, B, C$ lying inside the region and for other three materials sets lie relatively close to this region.
Such small deviations are  within the reported order of measurement accuracy ($0.25\%$, $0.3\%$ $0.6\%$ for Si, SiC and AlP, respectively). They can be eliminated by direct adjustments.
The fact that the Si belongs to that category in spite of its smaller band gap of $\approx 1.11$ can be easily explained. Indeed, it is one component diamond crystal with highly regular parabolic main valence and conduction bands diagrams, and additionally its structure follows the time reversal symmetry at $\Gamma$ point.

The rest of the materials from Table \ref{tabZBDRpar} have more complicated structure, e.g. the InSb is a small band gap, big effective mass material. It is known \cite{Kane1957}, that by accounting for the valence bands only, LK approximation would be insufficient for InSb like materials, and presented analysis support this fact theoretically. Concerning the Ge and GaAs they have anisotropic lower conduction bands without time reversal symmetry and high coupling between the $p$-bonding topmost valence band and $p$-antibonding conduction band states \cite{Pfeffer1990}.
Inclusion these conduction band states leads to more precise $8\times8$ and $14\times 14$ models \cite{Kane1957, Pfeffer1990}.
The setup described above is still applicable for these models but with a few minor modifications.

  \end{document}